\documentclass[12pt]{elsarticle}

\usepackage{amsthm}

\usepackage{soul}

\usepackage{graphicx}
\makeatletter
\def\set@curr@file#1{%
  \begingroup
    \escapechar\m@ne
    \xdef\@curr@file{\expandafter\string\csname #1\endcsname}%
  \endgroup
}
\def\quote@name#1{"\quote@@name#1\@gobble""}
\def\quote@@name#1"{#1\quote@@name}
\def\unquote@name#1{\quote@@name#1\@gobble"}
\makeatother
\usepackage{graphics}
\usepackage{epsfig}
\usepackage{float}
\usepackage{color,soul}
\usepackage{amsmath,amstext,amsbsy,amssymb}
\usepackage[cdot,amssymb]{SIunits}
\usepackage[latin1]{inputenc}
\usepackage{booktabs}
\usepackage{multirow}
\usepackage{array}
\usepackage{lineno}
\usepackage{setspace}
\usepackage{bm}
\usepackage{bbm}
\usepackage{amsmath,amssymb,amsfonts}
\newcommand{\x}{\mathbf x}
\newcommand{\f}{\mathbf f}

\begin{document}
\begin{frontmatter}



\title{Final draft of the paper https://doi.org/10.1016/j.ymssp.2020.107372 \\
 \textbf{A finite element model updating method based on global optimization}}


%
%

		\author[ISTI]{Maria Girardi}
		\ead{maria.girardi@isti.cnr.it}
		
		\author[ISTI]{Cristina Padovani}
		\ead{cristina.padovani@isti.cnr.it}
		
		\author[ISTI]{Daniele Pellegrini\corref{cor1}}
		\ead{daniele.pellegrini@isti.cnr.it}
		
		\author[ISTI,UNIPI]{Leonardo Robol}
		\ead{leonardo.robol@isti.cnr.it}
		
		\address[ISTI]{Institute of Information Science and
			    Technologies ``A. Faedo'', ISTI-CNR, Pisa, Italy.}
		
		\address[UNIPI]{Department of Mathematics, University of Pisa, Pisa, Italy.}
		
		    \cortext[cor1]{Corresponding author}

\begin{abstract}
Finite element model updating of a structure made of linear elastic materials is based on the solution of a minimization problem.
The goal is to find some unknown parameters of the finite element model (elastic moduli, mass densities, constraints and boundary conditions)
that minimize an objective function which evaluates the discrepancy between experimental and numerical dynamic properties.
The objective function depends nonlinearly on the parameters and may have multiple local minimum points. This paper presents a numerical method able
to find a global minimum point and assess its reliability. The numerical method has been tested on two simulated examples -- a masonry tower and a domed temple --
and validated via a generic genetic algorithm and a global sensitivity analysis tool. A real case study monitored under operational conditions has also been
addressed, and the structure's experimental modal properties have been used in the model updating procedure to estimate the mechanical properties of its constituent materials.

\end{abstract}

\begin{keyword}
Modal analysis, finite elements, model updating, global
optimization, sensitivity, masonry constructions



\end{keyword}

\end{frontmatter}

\section{Introduction}
\label{sec:introduction}

Finite element (FE) model updating is an essential component of numerical simulations in structural engineering \cite{douglas1982dynamic}, \cite{friswell2013finite}, \cite{marwala2010finite}. It aims to calibrate the FE model of a structure in order to match
numerical results with those obtained via experimental vibration tests. The calibration allows determining unknown structure's characteristics, such as
material properties, constraints, and boundary conditions.
While the main advantage of such calibration is an updated FE model that can be used to obtain more reliable predictions
regarding the dynamic behaviour of the structure, a further important application of model
updating is damage detection \cite{teughels2005damage}, \cite{ES2019}, \cite{giordano2019damage}.

FE model updating consists of solving a constrained minimum problem,  the objective function  being the distance between experimental and numerical quantities, such as the structure's natural
frequencies and mode shapes \cite{friswell2013finite}. Numerical modal properties depend on some unknown parameters, which
may suffer from a high degree of uncertainty mainly connected to the lack of information about both the structure's constituent materials and the interactions among its structural elements. In order to reduce the number of unknown parameters and make the minimum problem more manageable, it is possible to resort to sensitivity analysis \cite{Saltelli}, \cite{SAFE1},  \cite{Salib}, \cite{MSSP2011}, \cite{MSSP2019}, which allows assessing the influence of the parameters on the modal properties in order to exclude the less influential parameters from the model updating process.

Although application of FE model updating to historic masonry buildings is relatively recent, the literature on the subject is plentiful,
\cite{aoki2007structural},
\cite{gentile2007ambient},
\cite{ramos2010monitoring},
\cite{bayraktar2010finite},
\cite{ramos2011dynamic},
\cite{perez2011characterization},
\cite{araujo2012seismic},
\cite{costa2015updating},
 \cite{Boscato},
 \cite{ceravolo2016vibration},
\cite{cabboi2017continuous},
\cite{compan2017structural},
\cite{erdogan2017discrete},
\cite{fragonara2017dynamic},
\cite{kocaturk2017investigation},
\cite{torres2017operational},
\cite{altunicsik2018automated},
\cite{bassoli2018ambient},
\cite{bautista2018integrating},
\cite{ferraioli2018dynamic},
\cite{GenCha},
and  focused on case studies of historical interest for which a vibration-based
model updating is conducted. Preliminary FE models are calibrated using the modal properties determined
through system identification techniques. In the majority of the papers cited above  the FE modal analysis  is conducted
using commercial codes, and the model updating procedure is implemented separately.

Many papers have adopted a trial and
error approach (see, for example, \cite{costa2015updating}, \cite{bayraktar2010finite}), in which a manual fine-tuning procedure is used for
FE model updating. Such an approach is
impractical when the number of free parameters or the size of the model is large, in which case recourse to an automated model updating  becomes more advantageous.

The minimum problem stemming from FE model updating, whose objective function  may have  multiple local minima, can be solved via local or global minimisation procedures \cite{rios2013derivative}. The former may be based on trust-region schemes \cite{conn1988global}, while the latter rely on both deterministic and  stochastic  approaches, which encompass genetic, simulated annealing and particle swarm algorithms.

A deterministic approach to the optimisation using multi-start methods to avoid local minima has been proposed in \cite{GenCha}. In this work the global minimum point is selected from among several local minima calculated using different starting points chosen via the Latin Hypercube Sampling (LHS) method \cite{mckay1979comparison}.

A similar approach is adopted in \cite{teughels2005damage} and \cite{Bakir}, where the global optimization technique "Coupled Local Minimizers", based on pairwise state synchronization constraints, turns out to be more efficient than the multi-start local methods which rely on independent runs.

As far as sensitivity analysis is concerned, several parameter selection methods are available for choosing the unknown parameters  that should be considered in the FE model updating. Most are based on the matrix of local sensitivities, whose entries usually contain the partial derivatives of the numerical frequencies calculated  at a fixed parameter vector \cite{MSSP2011}.  Local sensitivity analysis (LSA) can only provide information about the behaviour of the frequencies in a neighbourhood of the given parameter vector and is  thus unable to provide any insight into the most relevant parameters influencing the frequencies. On the other hand, global sensitivity analysis (GSA)  \cite{Saltelli} provides a global measure of the dependence  of the frequencies on the parameters and represents a preliminary step in the model updating process, when the number and influence of the parameters are uncertain. Before tackling the optimization problem, it is worth mentioning, by way of example, the GSA applications described in  \cite{Boscato} and \cite{GenCha}. In particular,  in \cite{Boscato}  the results of a global sensitivity analysis based on the elementary effect (EE) method are compared with the results of a local sensitivity analysis, showing that the former performs better than the latter in model updating of the church of S. Maria del Suffragio in L'Aquila (Italy).   Instead, in  \cite{GenCha} an average sensitivity matrix is calculated via the LHS method, which is subsequently  adopted to calibrate
the Brivio bridge, a historic concrete structure in Lombardy, Italy.

A numerical method for solving the nonlinear least squares problem
involved in model updating  has been
proposed in \cite{JCAM} and \cite{FAC}. The algorithm, based on the
construction of local parametric reduced-order models embedded
in a trust-region scheme, was implemented in NOSA-ITACA, a
noncommercial FE code developed by the authors \cite{NOSA}, \cite{voltone}.
Similar approaches are described in \cite{AES2014} and \cite{GenCha}, where the numerical tools expressly developed for model updating are linked to commercial finite element codes used as a black-box within the framework of an iterative process. In particular, \cite{AES2014} presents  the MATLAB tool PARIS for automated FE model updating. PARIS is a research freeware code linked to the commercial software SAP2000, which has been applied to full-scale structures for damage detection purposes. The MATLAB procedure presented in  \cite{GenCha} relies instead on ABAQUS and its efficiency is tested on a historic concrete bridge.
Unlike the numerical procedures available in the literature, the algorithm
for solving the constrained minimum problem presented  in \cite{JCAM} and \cite{FAC} takes advantage of the fact
that the NOSA-ITACA source code is at the authors' disposal. This allows exploiting  the structure of the stiffness and mass matrices and the fact that only a few of the smallest eigenvalues have to be calculated. To compute these accurately, the natural choice is a (inverse) Lanczos method. When a parametric model is given, the
Lanczos projection can be interpreted as a parameter dependent model reduction, whereby only the relevant part of
the spectrum is matched. The Lanczos projection, combined  with a trust-region
method, allows matching the experimental frequencies with those predicted by the parametric model. This new procedure reduces  the overall computation
time of the numerical process and turns out to have excellent performance when compared to general-purpose optimizers. In addition, as the procedure described in \cite{JCAM} and \cite{FAC} allows calculating the singular value decomposition of the Jacobian of the residual function (the difference between experimental and numerical dynamic properties) at the minimum point, it makes it possible to assess the reliability of the parameters calculated and their sensitivity to noisy experimental dynamic properties.

In this paper, the numerical method proposed in \cite{JCAM} and \cite{FAC} to solve the constrained minimum problem encountered in FE model updating is modified in order to calculate a global minimum point of the objective function in the feasible set. This work is based on a deterministic approach, unlike the relatively recent large body of literature focused on stochastic model updating \cite{mares2006stochastic}, \cite{MSSP2019}, which aims to take into account and assess the uncertainties in both experimental data and numerical models as well.

Section 2 recalls the formulation of the optimization problem related to FE model updating. Then the global optimization method integrated into NOSA-ITACA is described, and some issues related to the reliability of the recovered solution are presented and discussed.  In particular, once the optimal parameter vector has been calculated, two quantities are introduced, which involve the partial derivatives of the numerical frequencies with respect to the parameters and provide a measure of how trustworthy the single parameter is.  Section 3 is devoted to testing the numerical method on two simulated examples: a masonry tower and a domed temple, which highlight  the capabilities and features of the global optimization algorithm proposed in Section 2.  For the sake of comparison, we also ran a global optimizer based on a genetic algorithm available in MATLAB. Such comparisons highlighted the excellent performance of the proposed method in terms of both computation time and number of evaluations of the objective function. Section 4 presents a real case study, the Matilde donjon in Livorno. This historic tower, which is part of the Fortezza Vecchia (Old Medici Fortress), was subjected to ambient vibration tests under operational conditions and its experimental dynamic properties used in the model updating procedure.

\section{The numerical method}
  \label{sec:numericalmethod}

The algorithms described in this section and used to perform
FE model updating through a global optimization
procedure are implemented in the NOSA-ITACA code
(www.nosaitaca.it). NOSA-ITACA code is
free software developed in
house by ISTI-CNR to disseminate the use of mathematical
models and numerical tools in the field of Cultural Heritage
\cite{voltone}. NOSA-ITACA combines
NOSA (the FE solver) with the graphic platform SALOME
(www.salome-platform.org) suitably modified and used to manage the
pre and post-processing operations. 
The code was developed to study the static and dynamic behaviour of
masonry structures \cite{TopPDD}, \cite{AG_SF}. To this end, it has been equipped with the
constitutive equation of \emph{masonry-like} materials, which models masonry as an isotropic nonlinear elastic
material with zero or weak tensile strength and infinite or bounded
compressive strength \cite{delpiero}, \cite{SPRING}.
In recent years, the code has been updated by adding several
features which now enable it to perform modal analysis \cite{porcelli2015solution},
\cite{SAHC1}, \cite{SAHC2}, \cite{azza2}, linear perturbation
analysis \cite{COMPDYN2017}, \cite{MAMS}, \cite{CONST} and model
updating \cite{JCAM}, \cite{FAC}, \cite{PSI}. The following subsection
\ref{subsec2-1} presents the  FE model calibration as a minimum problem and recalls the algorithm for model updating implemented
in NOSA--ITACA described in \cite{JCAM}  and \cite{FAC} (to which the reader is
referred for a detailed description). The
new features implemented in the code are explained in detail in subsections \ref{subsec2-2}, \ref{sec:sensitivity} and \ref{sec:Jacobianbased}.

\subsection{Finite element model updating as a minimization problem}\label{subsec2-1}
The term model updating refers to
a procedure aimed at calibrating a FE model in
order to match the experimental and numerical dynamic properties
(frequencies and mode shapes) of a structure. It is naturally
defined as an inverse
problem obtained from modal analysis, which in turn relies on the
solution of the generalized eigenvalue problem
\begin{equation}\label{gep_MK}
\mathbf{K} \mathbf{u} = \omega^2 \mathbf{M} \mathbf{u},
\end{equation}
where $\mathbf{K}$ and $\mathbf{M}\in \mathbb{R}^{n\times n}$ are
respectively the stiffness and mass matrices
of the structure discretized into
finite elements,  with  $n$ the total number of
degrees of freedom. Both $\mathbf{K}$ and $\mathbf{M}$ are usually sparse and banded, symmetric and positive definite. 
The eigenvalue $\omega^{2}_{i}$ is linked to
the structure's frequency $f_{i}$ by the relation
$f_{i}=\omega_{i}/(2\pi)$, and the eigenvector $\mathbf{u}^{(i)}$
represents the corresponding mode shape.
The model updating problem
can be formulated as an optimization problem by assuming that the
stiffness and mass matrices,  $\mathbf{K}$ and $\mathbf{M}$, are
functions of the parameter vector $\mathbf{x}$ containing the unknown characteristics of the structure (mechanical properties, mass densities, etc.),
\begin{equation}\label{param}
\mathbf{K} = \mathbf{K} (\mathbf{x}),  \textup{ \ \ }\mathbf{M}= \mathbf{M}
(\mathbf{x}),  \textup{ \ \ }\mathbf{x}\in \Omega.
\end{equation}

The set $\Omega$ of valid choices for the parameters
is a $p$-dimensional box of $\mathbb{R}^p$

\begin{equation}\label{box}
\Omega= [a_1,b_1]\times[a_2,b_2]...\times[a_p,b_p],
\end{equation}
for certain values ${a_i}<{b_i}$ for $i$=$1$....$p$. By taking
(\ref{param}) into account, equation (\ref{gep_MK}) becomes

\begin{equation}\label{gep_MKx}
\mathbf{K}(\mathbf{x}) \mathbf{u}(\mathbf{x}) = \omega(\mathbf{x})^2
\mathbf{M} (\mathbf{x}) \mathbf{u}(\mathbf{x}).
\end{equation}

The ultimate goal is to determine the optimal value of $\mathbf{x}$
that minimizes the objective function
$\phi(\mathbf{x})$ defined by
\begin{equation}\label{objfun}
\phi(\mathbf{x})= \displaystyle\sum_{i=1}^q w^{2}_i
[f_i(\mathbf{x})-\widehat{f}_i]^2
\end{equation}
within box $\Omega$.

The objective function involves the frequencies
and therefore depends nonlinearly on $\mathbf x$.
We denote by $\widehat{\mathbf{f}}$ the vector of the $q$
experimental frequencies to match, and by  $\mathbf{f}(\mathbf{x})=\frac{1}{2\pi}\sqrt{\bm\Lambda(\mathbf{x})}$ the vector of the numerical frequencies,
with $\bm\Lambda(\mathbf{x})$ being the vector containing the smallest $q$ eigenvalues of Eq. (\ref{gep_MKx}),
increasingly ordered
according to their magnitude.
%
%
The number $p$ of parameters to be
optimized is expected 
to be no greater than $q$. The vector $\mathbf{w}$ in Eq. (\ref{objfun}) encodes the weight that should be given to
each frequency in the optimization scheme.
If the goal is to minimize the distance between the vectors of the measured and computed frequencies in the usual Euclidean
norm, $w_i=1$, should be chosen. If, instead, relative accuracy on the frequencies is
desired, $w_i=\widehat{f}_i^{-1}$ is a natural choice. If some frequencies are to be ignored, it is
possible to set the corresponding component of \textbf{w} to zero. To keep the scaling uniform, the weight
vector is always normalized in order to have its norm equal to 1.

A numerical method to find a local minimum point of the objective function $\phi(\mathbf{x})$, which may have several local minima in set $\Omega$, is proposed in \cite{JCAM} and \cite{FAC}, where the authors  describe a new algorithm based on construction of local parametric reduced-order models embedded in a trust-region scheme, along with its implementation into the FE code NOSA-ITACA.
When the FE model depends on parameters, as in Eq. (\ref{gep_MKx}), and the number $n$ of degrees of freedom is very large, it is convenient to build  small-sized, reduced models able to efficiently approximate the behaviour of the original model for all parameter values.  Such reduced models have been obtained in \cite{JCAM} and \cite{FAC} through modification of the Lanczos projection scheme used to compute
the first eigenvalues and eigenvectors in Eq. (\ref{gep_MKx}) and to create a local model of objective function (\ref{objfun}) that is not costly to
evaluate and is at least first-order accurate. This local model is then used
in the region in which it is accurate enough to provide useful
information on the descent directions; this can be guaranteed by
suitably resizing the trust region, if necessary. It has been be proved
that, when the local models are accurate, convergence to a local
minimizer is guaranteed.

\subsection{Searching for global minima}\label{subsec2-2}

Several approaches can be adopted to minimize the objective function (\ref{objfun}) in the feasible set $\Omega$. They can be summarized as follows, ordered by increasing difficulty:
\begin{enumerate}
	\item Find a local minimum point of the objective function in $\Omega$.
	\item Search for the global minimum point of the objective function in $\Omega$.
	\item Identify all the local minimum points in $\Omega$ and hence, by assuming they are isolated,
recover the global minimum as well.
\end{enumerate}

In engineering applications the third approach is the most desirable. Not only does it
guarantee discovering the most "likely'' parameters, but also provides other values that
might be equally acceptable in terms of matching the structure's frequencies.
Engineering judgment, something complicated
to insert into an objective function, will then guide the choice of the most likely
parameter values.
In practice, the first approach is  easier and also computationally
less demanding than both the others, so it is often opted for.



Herein we propose a heuristic
strategy to improve the globalization property
of the method introduced in \cite{JCAM} and recalled in the preceding subsection. The goal is to improve the robustness of the method, while partially addressing approaches 2 and 3, without increasing  the computational cost excessively.
Due to the heuristic nature of the method, from a theoretical point of view, it is impossible
to guarantee that all the local minima will be found, but the effectiveness and robustness
of the method can be demonstrated through a few practical examples, which are described in the next section.


The proposed algorithm implemented in NOSA--ITACA code can be summarized in the following steps:
\begin{enumerate}[(a)]
	\item A local minimum is calculated on the original feasible set
	  $\Omega = [a_1, b_1] \times \ldots \times [a_p, b_p]$, using the method
	  from \cite{JCAM} and assuming the mid-point of $\Omega$ as starting point .
	  	  	  	  	
	\item For $j=1,...,p$, let us define $m_j = \frac{1}{2}(a_j + b_j)$ and
	  decompose the box $\Omega$ into the union of $2^p$ sets of the type
	
\begin{equation}\label{omegabar}
	  \bar \Omega = I_1 \times \ldots \times I_p
\end{equation}
	  with
\begin{equation}\label{Ij}
	   I_j \in \{[a_j,m_j], [m_j,b_j]\}, \textup{ \ \ }j=1,...,p.	
\end{equation}	
	   \item A local minimum point is then calculated on each of the subsets defined above (which have disjoint inner parts), starting at their mid-points.
	   If in all the subproblems, the minima
	  coincide with that calculated at step (a), or are
	  on the boundary, then the method stops.
	  Otherwise, the recursion
	  continues on the subsets where new local minima have been identified by following the process described in step (b).
\end{enumerate}

The method proposed here can run into difficulties when considering a large number
of parameters, as the number of subproblems to solve grows exponentially. However,
the following numerical experiments will show that it is still feasible for several cases
of interest.

Multi-start optimization approaches are commonly used to find global minima, for example in \cite{GenCha} the starting points are determined via a Latin Hypercube Sampling method and a set of local minimum points found, among which the global minimum point is identified. The algorithm proposed here does not execute a fixed number of runs, one for each starting point, but is based on a recursive procedure, which stops according to a given criterion. Like multi-start methods, the proposed procedure provides a set of local minimum points, including the global one.

The steps laid out above omit one aspect that is rather subtle and requires careful treatment:
how to identify two minimum points.
When working in floating-point arithmetic, and using a
stopping criterion linked to a specified tolerance, two different approximations $\mathbf{x}_0$ and $\mathbf{x}_1$
can be obtained starting from two different values for the parameters, even in the case of a single
minimum point.
It is therefore essential to be able to distinguish situations in which these parameters represent two different
minimum points from when instead they are just small perturbations of the same minimum
point, as explained in detail in the following subsection.


\subsection{Recognizing the same minimum points and related sensitivity issues}
\label{sec:sensitivity}

This section is devoted to the open question posed in the foregoing,
that is, how to recognise when two minimum points ``coincide'', up to some tolerance.
To answer this question, it is necessary to specify this concept more clearly.
Before addressing this issue, it is worth recalling that the problem of minimizing function $\phi$ in set $\Omega$ is a particular inverse problem, as it aims to calculate the unknown parameters of
the FE model of the structure under examination
using measurements carried out on it. Analysing minimum points provides a measure of how reliably each parameter has been determined,
and can identify (at the first order) those parameters which only
weakly influence the numerical frequencies, and as such, cannot be reliably
determined by the inverse problem.

According to \eqref{objfun}  and neglecting vector $\mathbf w$ for the sake of simplicity,  the objective function under consideration has the form,
\begin{equation}
  \phi(\mathbf x) = \lVert \mathbf f(\mathbf x) - \widehat{\mathbf f} \rVert_2^2, \quad \text{with} \quad
  \mathbf f(\x) = \begin{bmatrix}
    f_1(\mathbf x) \\
    \vdots \\
    f_q(\mathbf x) \\
  \end{bmatrix}.
\end{equation}

 Let $\mathbf x_0$ be a local minimum point of the objective function
and assume, up to performing a parameter rescaling, that
$\mathbf x_0$ is the vector with all components equal to $1$.

Assuming that the objective function is sufficiently regular, the first-order conditions for $\mathbf x_0$ to be a local minimum point imply
$\nabla \phi(\mathbf x_0) = 0$, where $\nabla \phi(\mathbf x_0)$ is the Jacobian of $\phi(\mathbf x)$
at $\mathbf x=\mathbf x_0$.
However, in practical
situations vector $\mathbf f$ is known only approximately,
with a tolerance $\epsilon$, so it is possible to introduce a definition of \emph{pseudominimum set}
which is robust to perturbation.

Given $\mathbf x_0$ such that
$\nabla \phi(\x_0) = 0$,
we define the
$\epsilon$-pseudominimum set at $\mathbf x_0$ as follows
\begin{equation}
  \mathcal{P}_\epsilon(\phi, \mathbf x_0) = \left\{
    \mathbf x  \ | \ \exists \delta \mathbf f \in \mathbb{R}^q \ \text{with} \ \lVert\delta \mathbf f\rVert_2 \leq \epsilon, \
    \nabla \phi_{\mathbf \delta \f}(\mathbf x) = 0
  \right\},
\end{equation}
where
\begin{equation}
  \phi_{\mathbf \delta \f}(\mathbf x) = \lVert
    \mathbf f(\mathbf x) - \widehat{\mathbf f} - \delta \mathbf f
  \rVert_2^2,
\end{equation}
which is equivalent to considering the set of minimum points of the objective function for
close-by frequency configurations, which are acceptable given a certain
tolerance, $\epsilon$, chosen by the user.

In other words, given two local minimum points $\mathbf {x_0}$ and $\mathbf {x_1}$ calculated via the scheme described in
the foregoing, the two points actually represent the same
``numerical'' minimum if $\mathbf x_1 \in \mathcal P_\epsilon(\phi, \mathbf x_0)$.
Note that this relation is symmetric\footnote{It is however not transitive, so
	it does not define an equivalence relation.}, that
is, $\mathbf x_1 \in \mathcal P_\epsilon(\phi, \mathbf x_0) \iff \mathbf x_0 \in \mathcal P_\epsilon(\phi, \mathbf x_1)$, so this definition is consistent.


Considering that $\lVert \mathbf x_0 - \mathbf x_1 \rVert_2$ is expected to be small
and using a first--order expansion\footnote{The dependency of
	the eigenvalues on the parameters is analytic almost
	everywhere in the domain, hence the Taylor expansion
	performed here can be rigorously justified.} of function $\f(\x)$
around $\mathbf x_0$, make it possible to calculate $\mathcal P_\epsilon(\phi, \mathbf x_0)$

\begin{equation}\label{Pe}
  \mathcal P_{\epsilon}(\phi, \x_0) = \left\{
    \x \ | \ \exists \lVert \delta \f \rVert_2 \leq \epsilon, \
    \nabla \f(\x_0)^T \nabla\f(\x_0) (\x - \x_0) = \nabla\f(\x_0)^T \delta \f
  \right\},
\end{equation}

where $\nabla \f(\x_0)$ denotes the Jacobian of $\f(\x)$ at $\x = \x_0$.

Let $\mathbf U \mathbf{\Sigma} \mathbf V^T = \nabla \f(\x_0)^T$ be the singular value decomposition (SVD) of
$\nabla\f(\x_0)^T$. By virtue of the fact that $\delta \f$ is arbitrary, and the multiplication
by unitary matrices leaves the Euclidean norm unchanged, it is possible to rewrite the set in (\ref{Pe}) as
follows
\begin{equation} \label{eq:ellipse}
    \mathcal P_{\epsilon}(\phi, \x_0) = \left\{
  \x \ | \
    \lVert \mathbf{\Sigma} \mathbf U^T (\x - \x_0) \rVert_2 \leq \epsilon
  \right\}.
\end{equation}
A SVD can be compute with $\mathcal O(q^2 p)$ flops, assuming $q \geq p$,
and is therefore a negligible cost in the proposed algorithm. Note in particular that the cost of computing this set is
independent of $n$, the degrees of freedom in the FE model. Hence,
\eqref{eq:ellipse} is easily verifiable in practice, and has been implemented
as a test in the algorithm described in the foregoing. The algorithm returns the matrices $\mathbf{\Sigma}$ and $\mathbf U$, which can be
used to construct the ellipsoid $\mathcal P_\epsilon(\phi, \x_0)$, which describes,
at the first-order, the level of accuracy attained in the space of parameters. In addition, the SVD of the Jacobian
can be used to compute, for each parameter $x_j$, the quantities
$\zeta_j$ and $\eta_j$, as described in the next subsection.

\subsection{Assessing the quality of the parameters}\label{sec:Jacobianbased}
Generally, experimental frequencies may not be accurate, since they
are derived by analyzing measured data that may be contaminated by
environmental noise. Thus, when minimizing objective function (\ref{objfun}),
one has to ensure that the optimal parameters are well-defined and robust
to perturbations in the data $\widehat {\mathbf f}$.

This analysis is only relevant in a neighbourhood of the minimum
point: the behaviour of the objective function elsewhere does not
influence the conditioning of the optimization problem.


A complete description of the parameters space and the directions
where the problem is well- or ill-defined can be given by computing
the SVD of the Jacobian, as is widely referenced in the numerical optimization
literature and pointed out for the problem at hand in \cite{FAC}. Nevertheless, if the  dimension of the parameter space is greater than three,
giving a meaningful interpretation to these directions can be difficult; hence, we introduce two quantities which are easier to interpret
and convey the same information.

Let $\widehat{\mathbf x}$ be a local minimum point of the nonlinear objective function (\ref{objfun}).
We assume that function $\mathbf f(\mathbf x)$ has been properly scaled so that both $\widehat {\mathbf x}$ and
$\widehat {\mathbf f}$ are vectors of all ones, and we
replace $\mathbf f(\mathbf x)$ with its first-order expansion at $\x= \widehat{\mathbf x}$.
We may now define the following
parameters for each $j = 1, \ldots, p$
\begin{equation} \label{eq:etaepsilon}
    \zeta_j := \left\lVert \frac{\partial \mathbf f}{\partial x_j} \right\rVert_2.
    \qquad
    \eta_j :=
      \min_{\mathbf v \in \mathcal S_j} \left\lVert
        \frac{\partial \mathbf  f}{\partial \mathbf v}
      \right\rVert_2,
\end{equation}
where $\frac{\partial \mathbf  f}{\partial\mathbf v}$ denotes the directional
derivative, and set $\mathcal S_j$ is defined as follows
\begin{equation}\label{eq:Sj}
  \mathcal S_j := \left\{
   \begin{bmatrix}
  \mathbf v_1 \\
  1 \\
  \mathbf v_2
  \end{bmatrix} \in \mathbb R^p \ | \ \left\lVert \begin{bmatrix}
  \mathbf v_1 \\ \mathbf v_2
  \end{bmatrix} \right\lVert_2 \leq 1, \
  \mathbf v_1 \in \mathbb R^{j-1}, \
  \mathbf v_2 \in \mathbb R^{p-j} \right\}.
\end{equation}
Note that set $\mathcal S_j$ contains, in particular, the $j$-th
vector $\mathbf e_j$ of the canonical basis of $\mathbb R^p$
, and therefore it must hold that $\eta_j \leq \zeta_j$.
Intuitively, $\mathcal S_j$ is the set of directions where the $j-th$ parameter
is forced to change at ``unit speed'', while the others can
change at some other speed, but are still bounded in the Euclidean
norm by $1$. Taking the minimum of the directional derivatives in $ \mathcal S_j$
is equivalent to finding the direction in the parameter space
with the slowest growth of $\mathbf f(\mathbf x)$, in which parameter
$x_j$ is involved.

Hence, we can make the following remarks:
\begin{itemize}
	\item If $\eta_j$ is small (i.e., $\eta_j \ll 1$),
	  then there exists a direction in which $x_j$ is forced to change,
	  but $\mathbf f(\mathbf x)$ varies slowly; hence,
	  determination of $x_j$ might be subject to noise. If, on the other hand, $\eta_j \gg 0$, then its determination through
	  the optimization problem is robust to noise.
    \item If $\zeta_j$ is small, then when $x_j$ changes, the
      frequencies are nearly unaffected; hence, there
      is no information on $x_j$ that can be obtained by solving
      the optimization problem. On the other hand, if $\zeta_j$ is large, then it cannot be guaranteed that $x_j$ is not affected by
      noise, but there is at least one direction in the parameter space
      involving $x_j$ that can be reliably determined. 	
\end{itemize}

The direction mentioned above can be determined from the SVD of the Jacobian $\nabla\mathbf f(\widehat{\mathbf  x}) = \mathbf U\mathbf{\Sigma} \mathbf V^T$, as described
in \cite{FAC}. However, parameters $\zeta_j$ and $\eta_j$
are easier to read, and we have the following trichotomy:
\begin{enumerate}[(i)]
	\item $\eta_j \leq \zeta_j \ll 1$: parameter $x_j$ cannot
	  be reliably determined, as no information on it is encoded in the
	  optimization problem.
	\item $0 \ll \eta_j \leq \zeta_j$: parameter $x_j$ can be
	  reliably determined from the data, even if it is subject to noise.
	  The amount of noise that can be tolerated is bounded in norm
	  by $\eta_j$.
	\item $\eta_j \ll 1$, but $\zeta_j \gg 0$: there is some information
	  on parameter $x_j$ encoded in the problem, but the result will not be free of noise. To find the directions which can be
	  ``trusted'', one has to look at the right singular vectors corresponding to large singular values in the SVD of the Jacobian.
\end{enumerate}
It is immediately clear that $\zeta_j$ can be computed directly by taking
the norms of the columns of the Jacobian.
Computing $\eta_j$, on the other hand, requires some more effort.
Let us temporarily drop the requirement that
$\lVert [\mathbf v_1^T \quad \mathbf v_2^T] \rVert_2 < 1$ in \eqref{eq:Sj}.
Thus, the minimizer $\mathbf v$ can be found by solving an unconstrained
linear least square problem, and in particular we have
\begin{equation}\label{zetam}
\mathbf v = \begin{bmatrix}
 \mathbf v_1 \\ 1 \\ \mathbf v_2 \\
\end{bmatrix}, \quad \text{with} \quad
\begin{bmatrix}
\mathbf v_1 \\ \mathbf v_2
\end{bmatrix} =
-\nabla\mathbf f(\widehat{\mathbf  x})_j^\dagger
\nabla\mathbf f(\widehat{\mathbf  x})\mathbf e_j ,
\end{equation}
where $\nabla\mathbf f(\widehat{\mathbf  x})_j$ is the Jacobian without the $j$-th column,
and the symbol $\mbox{}^\dagger$ denotes the
Moore-Penrose pseudoinverse. If $\lVert [\mathbf v_1^T \quad \mathbf v_2^T] \rVert_2 $
is less than $1$, then $\mathbf v$ in (\ref{zetam}) is the
minimizer for the constrained problem in \eqref{eq:etaepsilon} as well. Otherwise, an explicit formula is not available and we use the orthogonal
projection of the computed $\mathbf v$
onto $\mathcal S_j$ as a starting point
and determine the solution by solving a constrained
nonlinear least square problem. For solution of this
problem, we rely on the SQP algorithm described in Chapter
18 of \cite{nocedal}.

\section{Application to simulated case studies}
\label{sec:simulatedexamples}

In order to test the method described in section \ref{sec:numericalmethod}, two
artificial examples have been proposed. In both cases, the structure's free parameters are assigned, and a preliminary numerical modal analysis is performed
to evaluate the corresponding frequencies and mode shapes. Subsequently, the numerical frequencies
are employed as input to the model updating procedure to recover the original parameters.
The first example highlights the ability of the NOSA--ITACA code to discover more minimum points as compared to a generic genetic algorithm used to solve the same problem, which is unable to find more than one point.
The second example shows some of the code's features, which can help users to choose the most suitable optimal parameters
characterized by the greatest reliability.

The tests, conducted with NOSA-ITACA and MATLAB R2018b, were run on a computer with an Intel Core i7-8700
running at 3.20 GHz, with 64GB of RAM clocked at 2133MHz.



The weight vector $\mathbf w$ is always chosen to be $w_i = \widehat f_i^{-1}$, which ensures relative accuracy of the recovered frequency.

\subsection{A masonry tower}\label{subsec3-1}

\begin{figure}[H]
\centering
\includegraphics[width=14cm,trim=0 0 0 10]{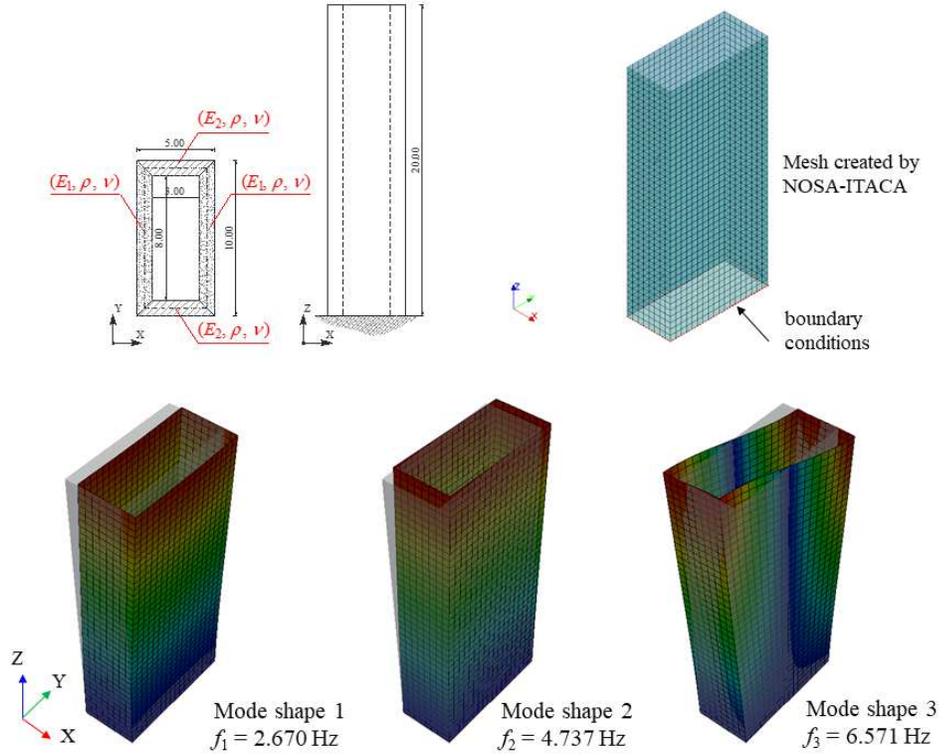}
\caption{The masonry tower: geometry (length in meters); model
created by NOSA-ITACA code; the first three mode shapes.}
\label{torre_num}
\end{figure}

As a first example, we considered the tower shown in Figure \ref{torre_num}. The $\unit{20}{\metre}$-high
structure has a rectangular cross section of $\unit{5}{\metre}\times\unit{10}{\metre}$
and walls of $\unit{1}{\metre}$ constant thickness. The tower, clamped at its base, is
discretized into $2080$ eight--node quadrilateral thin shell elements (element number $5$ of the NOSA-ITACA library \cite{NOSA}) for a total of $6344$ nodes
and $25376$ degrees of freedom.
A preliminary modal analysis is performed to evaluate the frequencies and mode shapes under the assumptions that the tower is made of a homogeneous material with Young's
moduli $E_1=E_2=3.00$ GPa (see Figure \ref{torre_num}), Poisson's ratio $\nu=0.2$ and mass density $\rho=\unit{1835.5}{\kilogram\per\metre^3}$.
The vector of the corresponding natural frequencies
obtained with the above parameters is

\begin{eqnarray}\label{freqarch}
\widehat{\mathbf{f}}=\unit{\text{[2.670, 4.737, 6.571]}}{\hertz}.
\end{eqnarray}

Figure \ref{torre_num} shows the mode shapes corresponding to the first three tower's frequencies:
the first two modes are bending movements along X and Y respectively, while the third is a torsional
mode shape.


The algorithm described in this paper is used to determine the Young's
moduli $E_1$ and $E_2$ of the structure. Putting $\mathbf{x}=[E_1, E_2]$, with the parameters varying within the interval

\begin{equation} \label{intarch}
\unit{1.00}{\giga\pascal}\leq\emph{$E_1,E_2$}\leq\unit{10.00}{\giga\pascal},
\end{equation}
model updating is conducted considering frequencies $\widehat f_1$ and $\widehat f_2$ in case (a), and $\widehat f_1$, $\widehat f_2$ and $\widehat f_3$ in case (b).

\begin{figure}[H]
\centering
\includegraphics[width=14cm,trim=0 0 0 10]{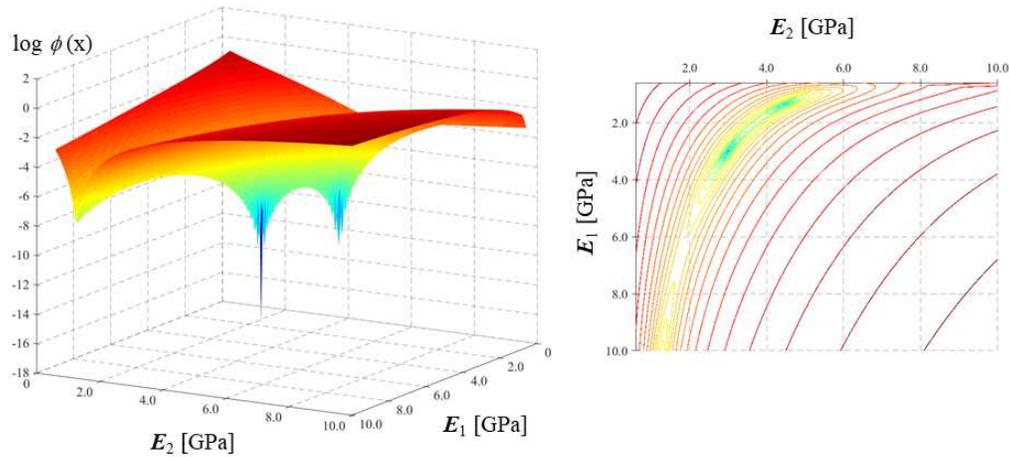}
\caption{Case (a) -- On the left a 3D plot of the objective function vs. $E_1$ and $E_2$. On the right a contour plot
of the same objective function where the two local minimum are clearly visible.}
\label{obj2freq}
\end{figure}

The same problems are also addressed with a generic genetic algorithm (denoted by GA) available in MATLAB R2018b, using NOSA--ITACA as a black box, with the aim of comparing the results of the two approaches and test the reliability and robustness of the numerical procedure proposed.
Table \ref{casea} summarizes the results related to case (a). Note firstly that
NOSA--ITACA code finds two minimum points, which correspond to the exact values of the known frequencies,
while the genetic algorithm calculates only one minimum, which is expected
be the global minimum point. The existence of two minimum points is shown in Figure \ref{obj2freq}, where
the plot of the objective function $\phi(\mathbf{x})$ defined in Eq. (\ref{objfun}) is reported in log--scale, as the two elastic moduli vary.
Regarding computation times and the number of evaluations of the objective function, the numerical procedure implemented in NOSA--ITACA appears to be much more efficient.

\begin{table}[H]\centering
\begin{tabular}[c]{c| c| c}
\toprule
 & $\text{NOSA--ITACA}$ & $\text{GA}$ \\
\hline
$\text{Minimum 1}$ & [3.00; 3.00] GPa & [3.02; 2.95] GPa\\
\hline
$\text{Frequencies}$ & [2.670, 4.737] Hz & [2.671, 4.732] Hz\\
\hline
$\text{Minimum 2}$ & [4.49; 1.34] GPa & --\\
\hline
$\text{Frequencies}$ & [2.670, 4.737] Hz & --\\
\hline
$\text{Computation time}$ & 11.50 s & 465.03 s\\
\hline
$\text{Number of evaluations}$ & 41 & 2600\\
\bottomrule
\end{tabular}
\caption{Case (a) -- Optimization results, two frequencies and two parameters.} \label{casea}
\end{table}

\begin{figure}[H]
\centering
\includegraphics[width=14cm,trim=0 0 0 10]{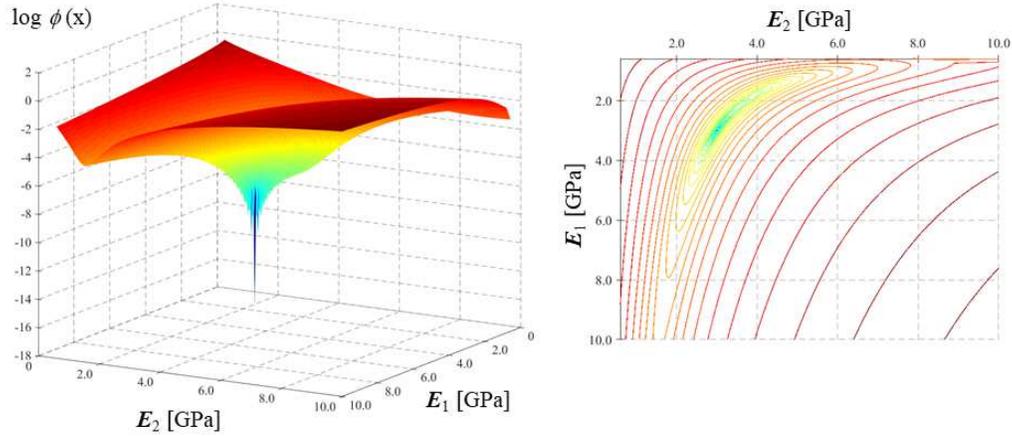}
\caption{Case (b) -- On the left a 3D plot of the objective function vs. $E_1$ and $E_2$. On the right a contour plot
of the same objective function where the only one local minimum is clearly depicted.}
\label{obj3freq}
\end{figure}

Regarding case (b), the results summarized in table \ref{caseb}
clearly show the superior performance of the NOSA--ITACA code in terms of both computation time and accuracy. Figure \ref{obj3freq} shows the plot of the objective function
$\phi(\mathbf{x})$, defined in Eq. (\ref{objfun}) and reported in log--scale,
which in this case exhibits one global minimum point.

\begin{table}[H]\centering
\begin{tabular}[c]{c| c| c}
\toprule
 & $\text{NOSA--ITACA}$ & $\text{GA}$\\
\hline
$\text{Minimum 1}$ & [3.00; 3.00] GPa & [3.00; 2.99] GPa\\
\hline
$\text{Frequencies}$ & [2.670, 4.737, 6.571] Hz & [2.670, 4.737, 6.571] Hz\\
\hline
$\text{Computation time}$ & 7.72 s & 497.63 s\\
\hline
$\text{Number of evaluations}$ & 27 & 2600\\
\bottomrule
\end{tabular}
\caption{Case (b) -- Optimization results, three frequencies and two parameters.} \label{caseb}
\end{table}

Table \ref{caseabJm} shows, for each minimum point
of cases (a) and (b), the parameters values $\zeta_j$ and $\eta_j$ defined
in subsection \ref{sec:Jacobianbased}. In all cases, $0 \ll \eta_j \ll \zeta_j$, which means
that every parameter $E_j$ has been determined reliably (as is evident in tables
\ref{casea} and \ref{caseb}) from the data, even if subject to noise.
The table also report $\zeta_j^{-1}$ and $\eta_j^{-1}$, quantities which provide an estimate
of the order of magnitude of the minimum and maximum percentage error (at the first-order) inherent in estimating the parameters
under the hypothesis of a 1\% error in the assessment of the experimental frequencies.
From the table it is clear that, in the worst-case scenario,
parameter estimation will be affected, at most, by a 6.2\% error in both cases (a) and (b).

\begin{table}[H]\centering
\begin{tabular}{c|c|c|c|c|c|c}
\toprule
 $\text{Case}$ & $\text{Minimum}$ & $x_j$ & $ \zeta_j $ & $ \eta_j $ & $ \zeta_j^{-1} $ & $ \eta_j^{-1} $ \\ \cline{1-7}
  \multirow{4}{*}{(a)} & \multirow{2}{*}{1} & $E_1$ & 1.0582 & 0.5061 & 0.945 & 1.976 \\
                     & & $E_2$ & 0.6001 & 0.1605 & 1.667 & 6.230\\ \cline{2-7}
                     & \multirow{2}{*}{2} & $E_1$ & 1.1257 & 0.6513 & 0.888 & 1.535 \\
                     & & $E_2$ & 0.5405 & 0.1946 & 1.850 & 5.138\\ \cline{1-7}
  \multirow{2}{*}{(b)} & \multirow{2}{*}{1} & $E_1$ & 1.2482 & 0.6255 & 0.801 & 1.598 \\
                     & & $E_2$ & 0.6630 & 0.1597 & 1.508 & 6.261\\
\bottomrule
\end{tabular}
\caption{Parameters $\zeta_j$ and $\eta_j$ for the cases (a) and (b).} \label{caseabJm}
\end{table}

\subsection{A domed temple}\label{subsec3-2}

\begin{figure}[H]
\centering
\includegraphics[width=14cm,trim=0 20 0 20]{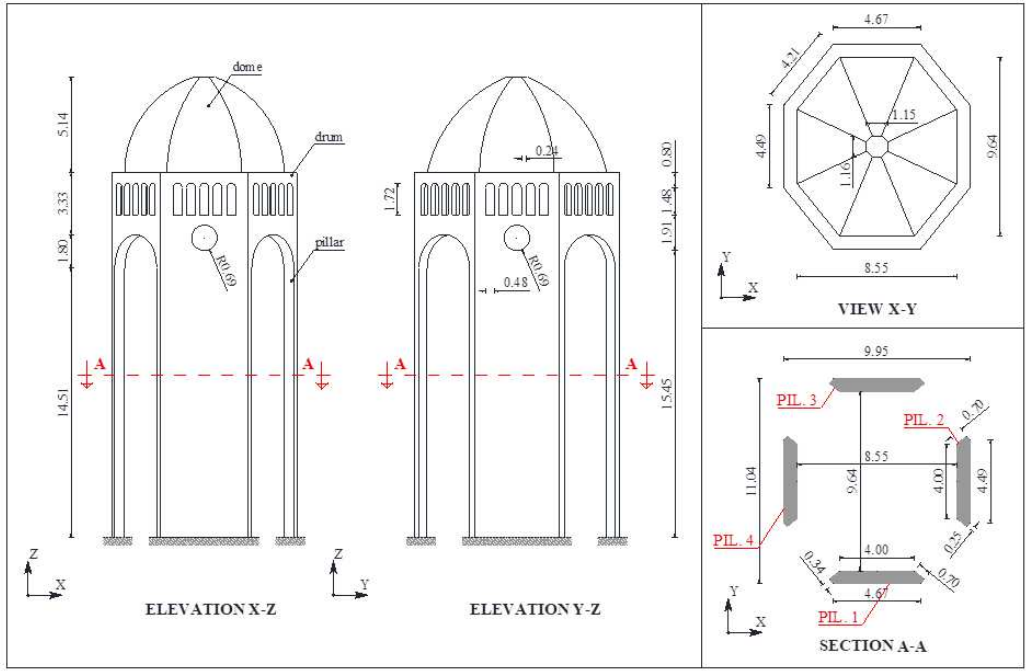}
\caption{Geometry of the domed temple (length in meters).}
\label{tempio}
\end{figure}

\begin{figure}[H]
\centering
\includegraphics[width=14cm,trim=0 20 0 20]{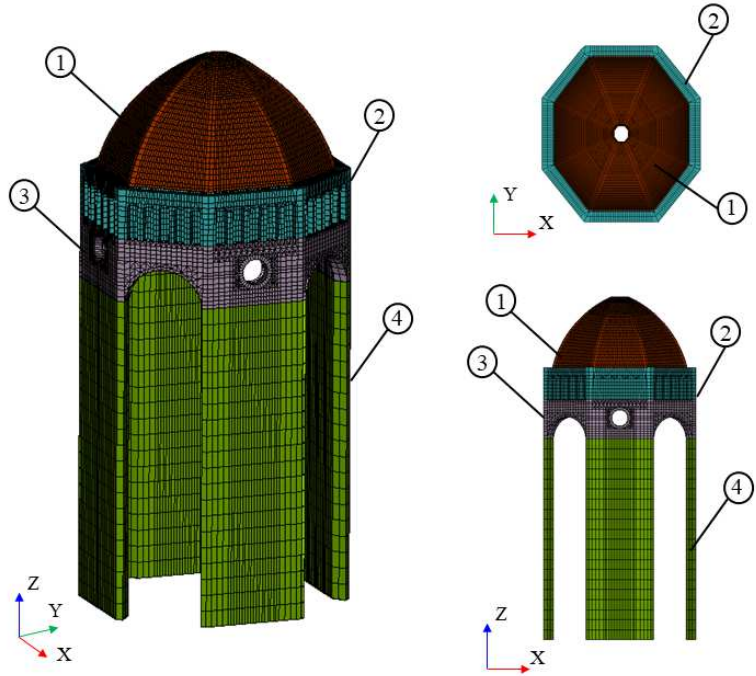}
\caption{Domed temple, mesh and materials. Each color corresponds to a different material, orange (1), cyan (2), violet (3) and green (4).}
\label{tempio_num}
\end{figure}

Let us now consider the domed temple, depicted in Figure \ref{tempio}, consisting of a $\unit{5}{\metre}$ high octagonal shaped cloister vault
resting on a drum inscribed on a
$\unit{10}{\metre}\times\unit{11}{\metre}$ rectangle. The structure, clamped at its base,
is made of $4$ different materials (Figure \ref{tempio_num}):
material $1$ for the dome (orange), material $2$ for the upper part of
the drum (cyan), material $3$ for the
bottom part of the drum (violet) and material $4$ for the columns (green).
The finite element model, shown in Figure \ref{tempio_num}, is composed of
$31052$ hexahedron brick elements
 and
$41245$ nodes for a total number of $123735$ degrees of freedom.

A preliminary modal analysis is performed to evaluate the
structure's frequencies assuming the material
properties reported in table \ref{tab_temple_mat}. The vector of the first eight natural frequencies is

\begin{eqnarray}\label{freqdome}
\widehat{\mathbf{f}}=\unit{\text{[2.19, 2.23, 3.76, 3.83, 4.32, 4.60, 4.72, 8.26]}}{\hertz}.
\end{eqnarray}

\begin{table}[H]\centering
\begin{tabular}[c]{c| c| c c c}
\toprule
$\text{Material}$ & $\text{Temple portion}$ & $\rho [\text{kg/m\textsuperscript{3}}]$ & $E [\text{GPa}]$ & $\nu $\\
\hline
1 (orange) & \text{dome} & 1800.0 & 3.00 & 0.25\\
\hline
2 (violet) & \text{drum (top)} & 1900.0 & 3.50 & 0.25\\
\hline
3 (cyan) & \text{drum (bottom)} & 2000.0 & 4.00 & 0.25\\
\hline
4 (green) & \text{pillars} & 2200.0 & 5.00 & 0.25\\
\bottomrule
\end{tabular}
\caption{Values of the material properties.} \label{tab_temple_mat}
\end{table}

The optimization code implemented in NOSA--ITACA and a generic genetic algorithm
were run setting $\mathbf x = [E_1, \rho_1, E_2, E_3, \rho_3, E_4, \rho_4]$, with the following bounds

\begin{equation}
\unit{2.00}{\giga\pascal}\leq\emph{$E_j$}\leq\unit{10.00}{\giga\pascal},  \textup{ \ \ } j=1,...,4,
\end{equation}

\begin{equation}
\unit{1600.0}{\kilogram\per\cubic\metre}\leq\emph{$\rho_j$}\leq\unit{2400.0}{\kilogram\per\cubic\metre}, \textup{ \ \ } j=1, 3, 4.
\end{equation}
This choice leaves seven parameters to be optimized, with the sole exception of $\rho_2$, which was set to the fixed value reported in table \ref{tab_temple_mat}.
Tables \ref{templep} and \ref{templef} summarize the results obtained
by NOSA--ITACA code and the genetic algorithm in terms of optimal parameter values, frequencies, relative errors $|\Delta_{x_j}|$ and $|\Delta_f|$, computation time and
number of evaluations of the objective function.

\begin{table}[H]\centering
\begin{tabular}[c]{c| c| c c| c c}
\toprule
 & $\text{Real value}$ & $\text{NOSA--ITACA}$ & $|\Delta_{x_j}|[\%]$ & $\text{GA}$  & $|\Delta_{x_j}|[\%]$\\
\hline
$E_1 [\text{GPa}]$  & 3.000 & 2.996 & 0.13 & 4.1431 & 38.10\\
\hline
$\rho_1 [\text{kg/m\textsuperscript{3}}]$ & 1800.0 & 1908.9 & 6.05 & 1988.6 & 10.47\\
\hline
$E_2 [\text{GPa}]$ & 3.500 & 4.085 & 16.72 & 4.0335 & 15.24\\
\hline
$E_3 [\text{GPa}]$ & 4.000 & 4.177 & 4.43 & 3.8357 & 4.11\\
\hline
$\rho_3 [\text{kg/m\textsuperscript{3}}]$ & 2000.0 & 2115.9 & 5.80 & 2340.1 & 17.00\\
\hline
$E_4 [\text{GPa}]$ & 5.000 & 5.132 & 2.63 & 5.6213 & 12.43\\
\hline
$\rho_4 [\text{kg/m\textsuperscript{3}}]$ & 2200.0 & 2272.7 & 3.30 & 2397.8 & 9.00\\
\hline
Computation time [s]&   & 14019 &  & 103250 & \\
\hline
Number of evaluations&  & 671 &  & 10500 & \\
\bottomrule
\end{tabular}
\caption{Optimal parameter values calculated by NOSA--ITACA code and a genetic algorithm.} \label{templep}
\end{table}


\begin{table}[H]\centering
\begin{tabular}[c]{c| c| c c| c c}
\toprule
 & $\text{Real value}$ & $\text{NOSA--ITACA}$ & $|\Delta_{f}|[\%]$ & $\text{GA}$  & $|\Delta_{f}|[\%]$\\
\hline
$f_1 [\text{Hz}]$ & 2.19 & 2.18 & 0.46 & 2.18 & 0.46\\
\hline
$f_2 [\text{Hz}]$ & 2.23 & 2.22 & 0.45 & 2.22 & 0.45\\
\hline
$f_3 [\text{Hz}]$ & 3.76 & 3.75 & 0.27 & 3.77 & 0.27\\
\hline
$f_4 [\text{Hz}]$ & 3.83 & 3.83 & 0.00 & 3.83 & 0.00\\
\hline
$f_5 [\text{Hz}]$ & 4.32 & 4.31 & 0.23 & 4.31 & 0.23\\
\hline
$f_6 [\text{Hz}]$ & 4.60 & 4.60 & 0.00 & 4.61 & 0.22\\
\hline
$f_7 [\text{Hz}]$ & 4.72 & 4.72 & 0.00 & 4.72 & 0.00\\
\hline
$f_8 [\text{Hz}]$ & 8.26 & 8.25 & 0.12 & 8.24 & 0.24\\
\bottomrule
\end{tabular}
\caption{Frequencies values corresponding to the parameters' optimal values recovered
 by NOSA--ITACA code and a genetic algorithm.} \label{templef}
\end{table}

The results above highlight that: (i) the numerical procedure implemented in NOSA--ITACA is less time--consuming than the genetic algorithm, the computation time of the former being ten times lower than that of the latter; (ii) the optimal values of the Young's moduli calculated by NOSA--ITACA are affected by a maximum relative error of $17$\%, against $38$\% of the genetic algorithm; (iii) the maximum relative error on mass density is about $6$\% for NOSA--ITACA and $17$\% for the genetic algorithm; (iv) even though the optimal value of some mechanical characteristics is affected by high error, the maximum relative error on the frequencies is about $0.5$\% for both numerical methods.

To investigate the robustness and reliability of the solution found,
the parameters values $\zeta_j$ and $\eta_j$ defined
in subsection \ref{sec:Jacobianbased} are reported in table \ref{Jmtemple} with their respective inverse values and the relative error $|\Delta_{x_j}|$
calculated in table \ref{templep}.

\begin{table}[H]\centering
\begin{tabular}[c]{c| c c| c c|c}
\toprule
 & $ \zeta_j $ & $ \eta_j $ & $ \zeta_j^{-1} $ & $ \eta_j^{-1} $ & $|\Delta_{x_j}|[\%]$\\
\hline
$E_1$  & 5.8216$\cdot10^{-2}$ & 2.4242$\cdot10^{-2}$ & 17.177 & 41.250 & 0.13\\
\hline
$\rho_1$  & 1.7265$\cdot10^{-1}$ & 1.0859$\cdot10^{-1}$ & 5.792 & 9.209 & 6.05 \\
\hline
$E_2$  & 7.4616$\cdot10^{-2}$ & 2.6615$\cdot10^{-2}$  & 13.402  & 37.573 & 16.72 \\
\hline
$E_3$  & 3.5101$\cdot10^{-1}$ & 2.4958$\cdot10^{-1}$  & 2.849  & 4.007 & 4.43 \\
\hline
$\rho_3$  & 3.3679$\cdot10^{-1}$ & 1.6885$\cdot10^{-1}$  & 2.969  & 5.922 & 5.80 \\
\hline
$E_4$  & 1.2272 & 9.2428$\cdot10^{-1}$  & 0.815  & 1.082 & 2.63 \\
\hline
$\rho_4$  & 1.1730 & 8.6633$\cdot10^{-1}$ & 0.853  & 1.154 & 3.30 \\
\bottomrule
\end{tabular}
\caption{Parameters $\zeta_j$ and $\eta_j$ calculated by NOSA--ITACA.} \label{Jmtemple}
\end{table}

The above table shows that the Young's moduli of materials $1$ and $2$ (the dome and the upper part of the drum) seem to be irrelevant in the optimization process.
This fact can be explained by observing the mode shapes related to the first eight frequencies, which mainly involve displacement of the pillars.
It is also interesting to note that the objective function is more heavily influenced by the dome's mass density than by
its elastic modulus ($\zeta_1=5.8216\cdot10^{-2}$ versus $\zeta_2=1.7265\cdot10^{-1}$), in line with the fact that the dynamic behavior of the structure is comparable
to a cantilever beam with a mass concentrated at the free end. The Young's
moduli and mass density of materials $3$ and $4$
seem more reliable than the others, as shown by the values of $\zeta_j$ and $\eta_j$.
Finally, note that the relative error $|\Delta_{x_j}|$ made in estimating the optimal values of the parameters is always close
to the range defined by $\zeta_j^{-1}$ and $\eta_j^{-1}$ (at the first-order, under the hypothesis of a maximum error of $1\%$
in the assessment of the experimental frequencies).

Further information can be achieved by calculating, at the minimum point, the scaled Jacobian matrix described
in subsection \ref{sec:Jacobianbased},
\begin{equation}\label{Jmat}
  \begin{pmatrix}
    7.32\cdot10^{-3} & -9.34\cdot10^{-2} & 2.61\cdot10^{-2} & 1.09\cdot10^{-1} & -1.23\cdot10^{-1} & 3.57\cdot10^{-1} & -1.77\cdot10^{-1} \\
    6.93\cdot10^{-3} & -9.05\cdot10^{-2} & 2.70\cdot10^{-2} & 1.07\cdot10^{-1} & -1.23\cdot10^{-1} & 3.60\cdot10^{-1} & -1.81\cdot10^{-1} \\
    1.03\cdot10^{-2} & -7.88\cdot10^{-4} & 2.02\cdot10^{-2} & 8.53\cdot10^{-2} & -2.74\cdot10^{-2} & 3.84\cdot10^{-1} & -4.66\cdot10^{-1}  \\
    1.03\cdot10^{-2} & -4.82\cdot10^{-2} & 2.01\cdot10^{-2} & 9.77\cdot10^{-2} & -1.53\cdot10^{-1} & 3.75\cdot10^{-1} & -1.77\cdot10^{-1}  \\
    6.15\cdot10^{-4} & -6.26\cdot10^{-5} & 1.32\cdot10^{-2} & 1.12\cdot10^{-1} & -3.15\cdot10^{-3} & 3.74\cdot10^{-1} & -4.97\cdot10^{-1}  \\
    1.58\cdot10^{-3} & -3.13\cdot10^{-2} & 1.39\cdot10^{-2} & 1.02\cdot10^{-1} & -2.61\cdot10^{-2} & 3.83\cdot10^{-1} & -4.10\cdot10^{-1}  \\
    1.05\cdot10^{-3} & -2.85\cdot10^{-2} & 1.03\cdot10^{-2} & 1.06\cdot10^{-1} & -2.65\cdot10^{-2} & 3.83\cdot10^{-1} & -4.14\cdot10^{-1}  \\
    4.63\cdot10^{-2} & -9.64\cdot10^{-3} & 3.54\cdot10^{-2} & 1.15\cdot10^{-1} & -1.57\cdot10^{-1} &3.04\cdot10^{-1}  & -2.78\cdot10^{-1}
  \end{pmatrix}
\end{equation}



The numbers reported in the first three columns of the matrix confirms that the temple's frequencies
are weakly dependent on materials $1$ and $2$. Restricting the attention to the last two columns
in matrix (\ref{Jmat}) (containing the partial derivatives of the frequencies with respect to $E_4$ and $\rho_4$) furnishes more information about the minimum point.
The SVD of the restricted matrix yields the results summarized in table \ref{SVDtemple}, with the singular values $\sigma_1 > \sigma_2$ reported in the first columns,
and the corresponding right singular vectors in the second and third columns. The objective function is expected to have  a direction with a weaker influence on the frequencies parallel to $\mathbf z^{(2)}$
(with constant ratio $E_4$/$\rho_4$), which corresponds to the smallest singular value $\sigma_2=2.5063\cdot10^{-1}$.

\begin{table}[H]\centering
\begin{tabular}[c]{c| c c}
\toprule
 $ \sigma $ & $ \mathbf z^{(1)} $ & $ \mathbf z^{(2)}$\\
\hline
1.4087 & -7.2408$\cdot10^{-1}$ & -6.8971$\cdot10^{-1}$\\
\hline
2.5063$\cdot10^{-1}$ & 6.8971$\cdot10^{-1}$ & -7.2408$\cdot10^{-1}$\\
\bottomrule
\end{tabular}
\caption{Singular values and right singular vectors of the scaled restricted Jacobian matrix.} \label{SVDtemple}
\end{table}

To investigate how variation in the input (Young's moduli and the mass densities of the domed temple's four constituent materials) influence the output of the numerical model (the natural frequencies),
and thereby test the sensitivity analysis implemented in the NOSA--ITACA code, a Global Sensitivity Analysis (GSA)
has been performed through the SAFE Toolbox \cite{SAFE1}, \cite{SAFE2} and \cite{SAFE3}.

\begin{figure}[H]
\centering
\includegraphics[width=14cm,trim=0 10 0 10]{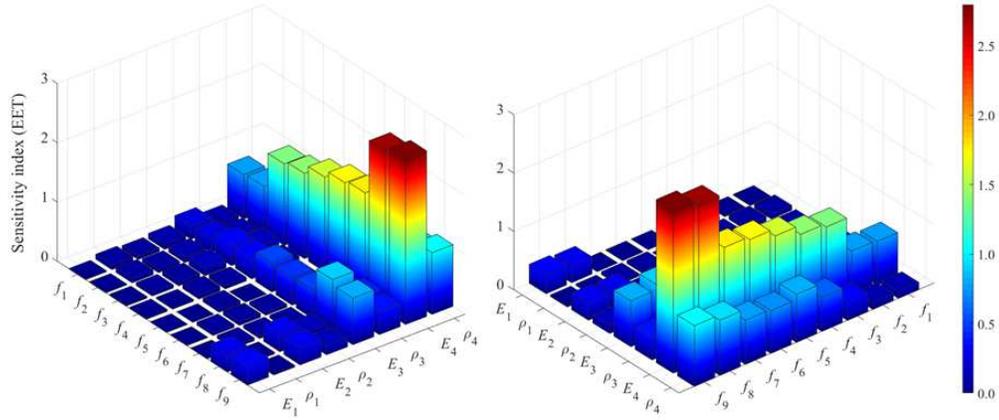}
\caption{EET sensitivity indices for the first nine frequencies and eight parameters.}
\label{EET_tempio}
\end{figure}

The SAFE Toolbox, an open--source code implemented in MATLAB,
can be easily linked to simulation models running outside the MATLAB environment, such as the NOSA-ITACA code in the example at hand.
The Elementary Effects Test (EET method \cite{Morris})
is used to evaluate the sensitivity indices assuming that the eight input parameters (Young's moduli and the mass densities of the four materials) have
a uniform probability distribution function, and adopting the Latin Hypercube method \cite{mckay1979comparison} as sampling
strategy.
From Figure \ref{EET_tempio}, where the sensitivity indices calculated
via the EET method are plotted, it is possible to deduce that the  Young's moduli of materials 3 and 4 affect the numerical frequencies much more than
the remaining parameters.
These results confirm the information recovered by the quantities $\zeta_j$ and $\eta_j$ calculated by NOSA--ITACA and reported in table \ref{Jmtemple}.

Sensitivity analysis, similar to the one reported in Figure \ref{EET_tempio}, is generally performed to choose the number of updating parameters and to exclude some uncertain parameters from the model updating process. It is interesting to observe that the results confirm the information obtained on the quality of the optimal parameters.
It is also worth noting that the computational cost of such a global sensitivity analysis is very high (Figures \ref{EET_tempio} is the results of $1260$ FE modal analysis runs) with respect to the cost of the minimization procedure implemented in NOSA-ITACA, which provides both the global minimum point and an assessment of its reliability.

\section{Application to a real example: the Matilde donjon in Livorno}
\label{sec:mastio}

\subsection{Experimental tests and dynamic identification}\label{subsec4-1}

\begin{figure}[H]
\centering
\includegraphics[width=14cm,trim=0 80 0 80]{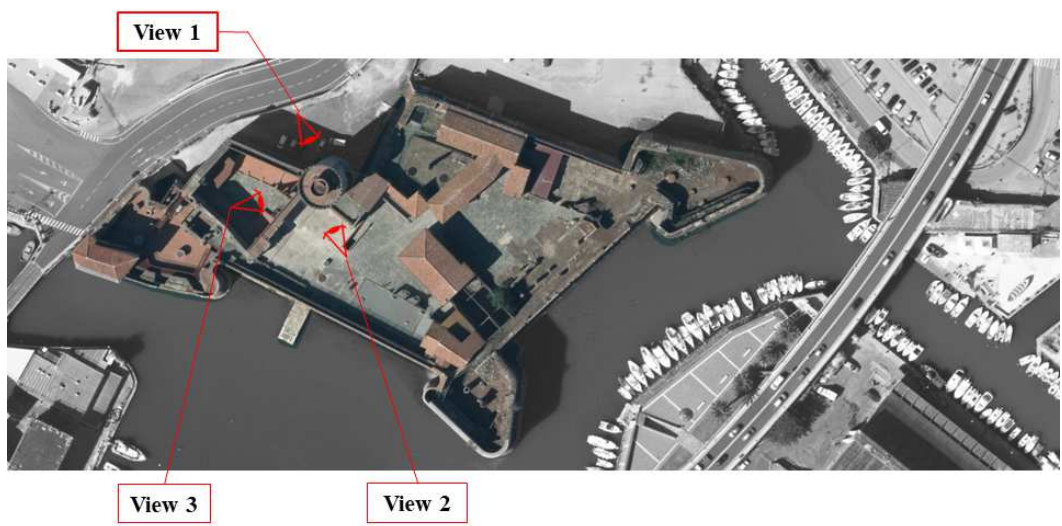}
\caption{The ``Old Fortress''(photo taken from
www.livornoportcenter.it).} \label{fortezzavecchia}
\end{figure}

The Matilde donjon is a fortified keep belonging to the Fortezza Vecchia (Old
Fortress), near the ancient Medici Port of Livorno, Italy (Figure
\ref{fortezzavecchia}).

\begin{figure}[H]
\centering
\includegraphics[width=14cm,trim=0 30 0 30]{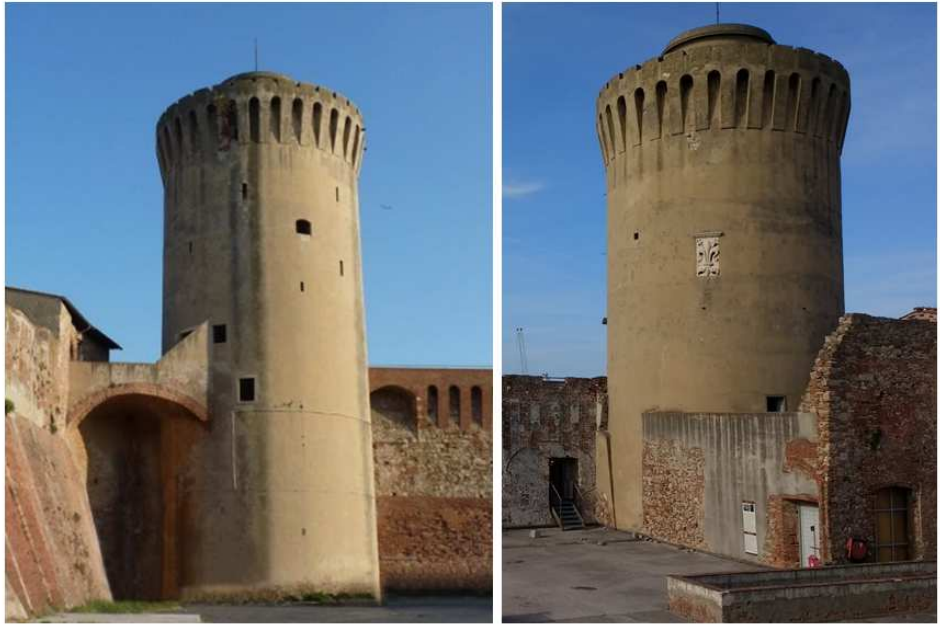}
\caption{The Matilde donjon (view 1, 2).}
\label{mastio_pw12}
\end{figure}

\begin{figure}[H]
\centering
\includegraphics[width=14cm,trim=0 40 0 40]{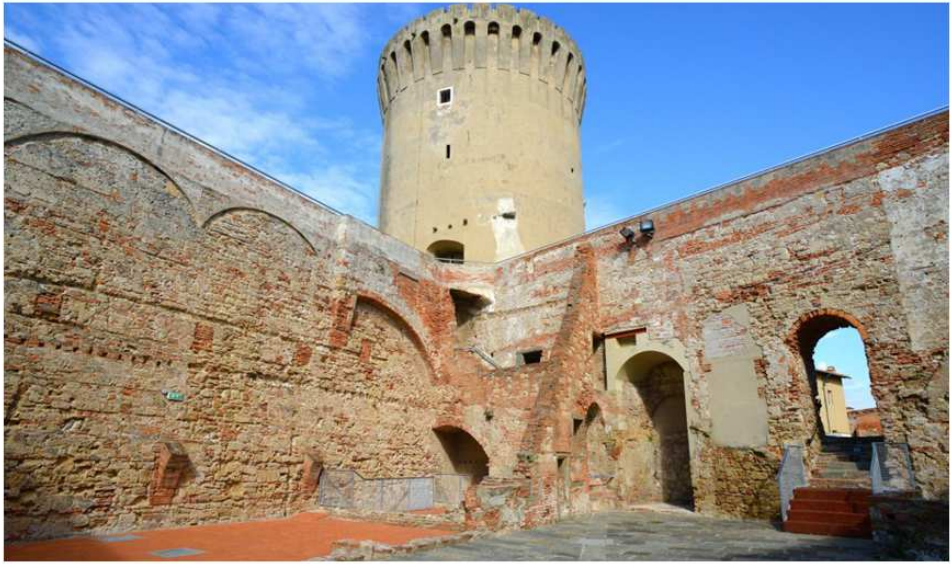}
\caption{The Matilde donjon (view 3).} \label{mastio_pw3}
\end{figure}

\begin{figure}[H]
\centering
\includegraphics[width=14cm,trim=0 20 0 20]{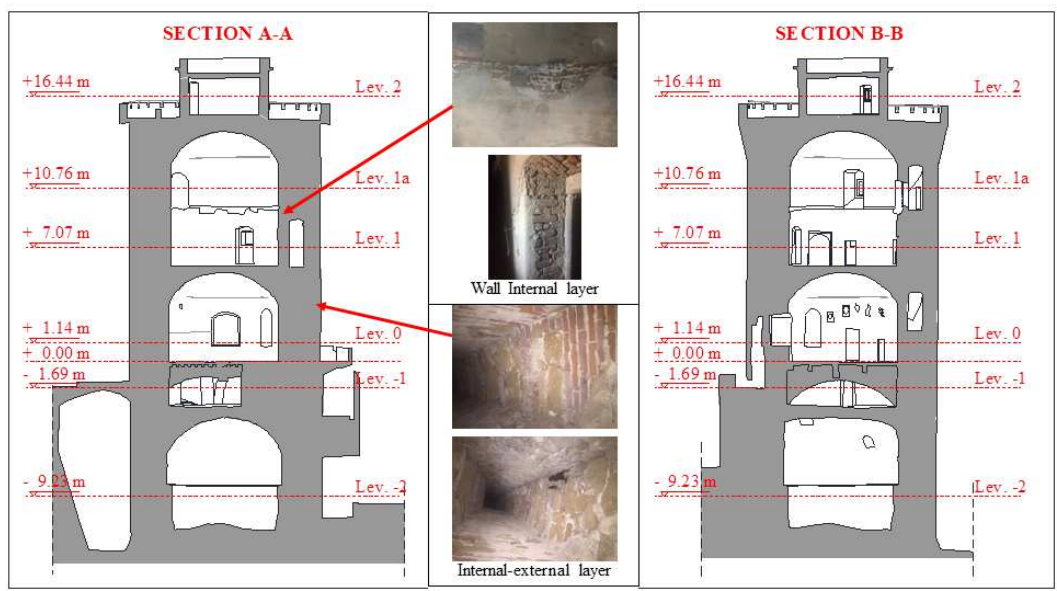}
\caption{Transverse sections of the tower.} \label{sez_Mastio}
\end{figure}

The $\unit{26}{\metre}$--high cylindrical tower shown in Figures \ref{mastio_pw12} and \ref{mastio_pw3} has a cross-section with a mean outer radius
of $\unit{6}{\metre}$ and walls of $\unit{2.5}{\metre}$ constant thickness along height \cite{barsocchi2020wireless}.
Although no precise information is available on its mechanical properties of the constituent materials, by visual inspection
the tower appears to be made of mixed brick-stone masonry with an internal layer made of clay bricks and mortar joints,
and the outer, more irregular layer of stone blocks and bricks. The tower's interior hosts four vaulted rooms (Figure \ref{sez_Mastio}).
At its base there is a large cistern, about $\unit{6}{\metre}$ high, for collecting rainwater.
A helicoidal staircase is found within the tower's wall, starting from the so-called ``Captains'' room at
level $0$ (see section Figure \ref{sez_Mastio}) and allows reaching the upper floor and the roof terrace,
crowned by cantilevered merlons.
The tower is tightly connected to the Old Fortress' external walls for a height of about $\unit{9}{\metre}$
from the level of the lower galleries (see Figures \ref{mastio_pw12} and \ref{mastio_pw3}).

\begin{figure}[H]
\centering
\includegraphics[width=14cm,trim=0 10 0 10]{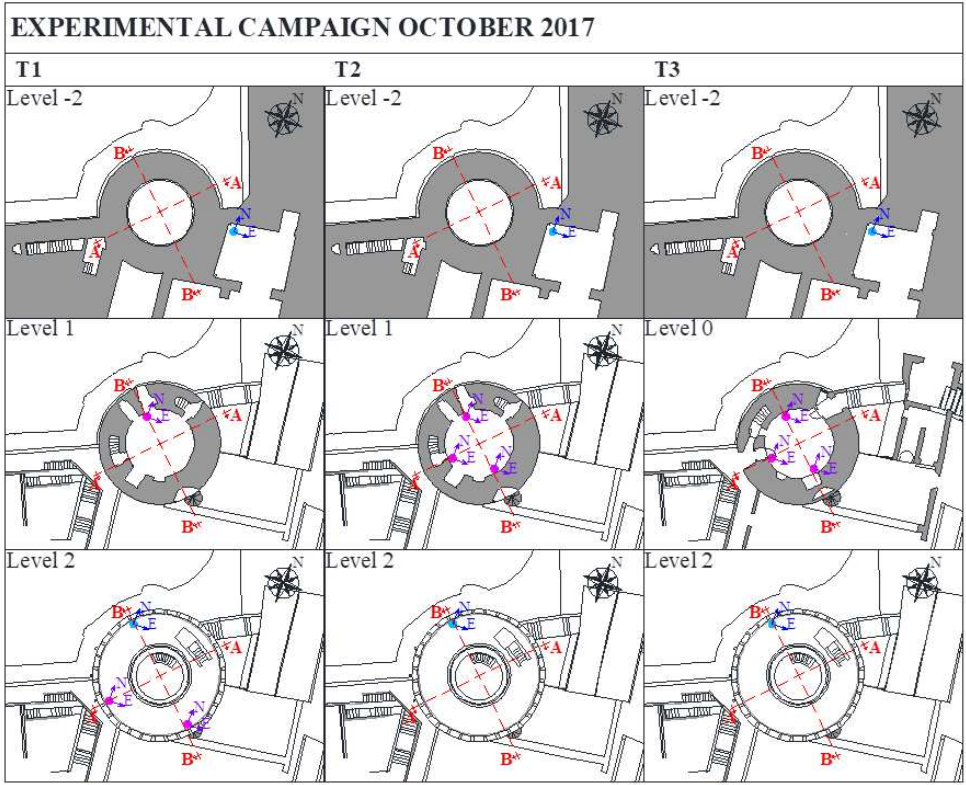}
\caption{Sensor layout October 2017 -- test T1, T2, T3.} \label{Mastio_Test123}
\end{figure}

\begin{figure}[H]
\centering
\includegraphics[width=14cm,trim=0 10 0 10]{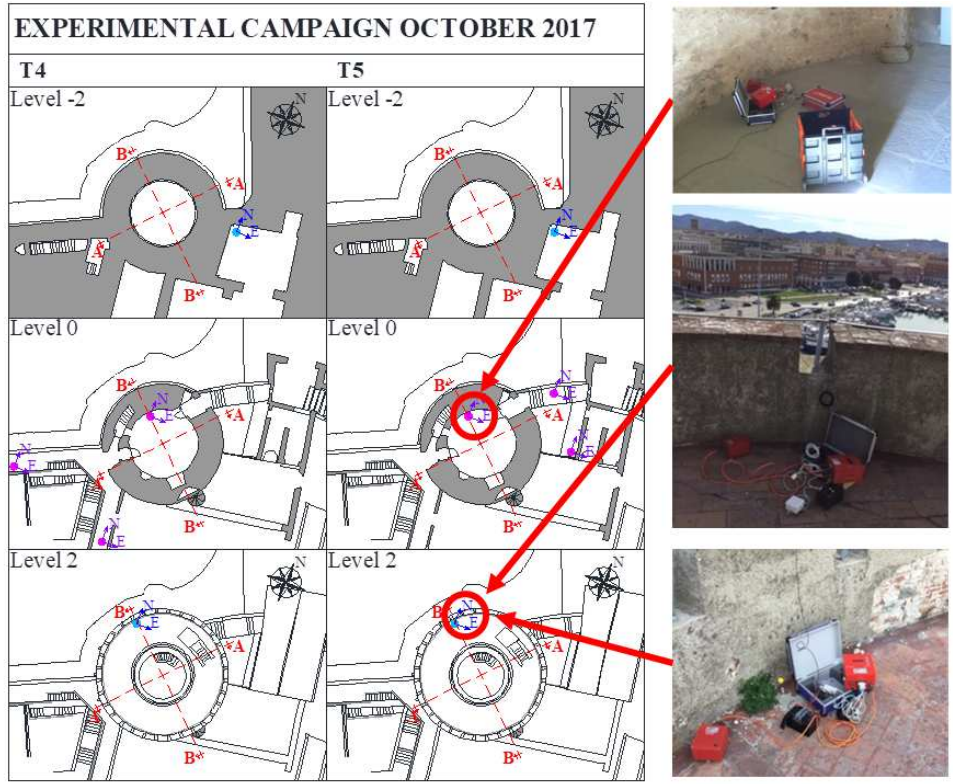}
\caption{Sensor layout October 2017 -- test T4, T5.} \label{Mastio_Test45}
\end{figure}

In October $2017$, an ambient vibration monitoring experiment was carried out on the tower (see Figure \ref{sez_Mastio},
\ref{Mastio_Test123}, \ref{Mastio_Test45}).
The ambient vibrations were monitored for a few hours via SARA SS$20$ seismometric stations
(https://www.sara.pg.it/) arranged in different layouts. During
the five tests (T$1$ to T$5$), each lasting about thirty
minutes, two sensors were kept in a fixed position-- one at the base
(level -$2$) and the other on the roof terrace (level $2$)-- while
the remaining sensors were moved to different positions along the
tower's height and surrounding area in order to obtain
information on the mode shapes and degree of connection between the
Old Fortress' structures and the tower itself. 
The sampling rate was set at \unit{100}{\hertz}.
All data recorded have been divided into short sequences, each lasting $1000$ seconds (a time window greater than the
structure's fundamental period estimated by preliminary FE modal analysis), and
processed by two different operational modal analysis (OMA) techniques,
through which the tower' modal parameters were estimated:
the Stochastic Subspace Identification covariance driven method (SSI--cov) \cite{peet2} implemented in MACEC code \cite{MACEC} and
the Enhanced Frequency Domain Decomposition method (EFDD) \cite{batel} implemented by ISTI--CNR in Trudi code \cite{pelle}.

In total, six vibration modes were identified in the frequency
range of $2$-$\unit{13}{\hertz}$. Table \ref{tab_freq_o} summarizes the results in terms of natural frequencies $f$, damping ratios $\xi$, and MAC (Modal
Assurance Criterion)\footnote{MAC is the scalar quantity which expresses the correlation between
two mode shapes, varying from from $0$ to $1$.} values
\cite{allemang1982correlation} calculated between the corresponding mode shapes
estimated via the two OMA techniques.

For the sake of brevity, the values shown in the tables correspond
to the average values of the estimated parameters during each test,
all of which are characterized by a MPC (Modal Phase Collinearity)\footnote{MPC is a parameter ranging from $0$ to $1$ that quantifies the
complexity of an eigenvector; MPC is $1$ for real vectors.} value \cite{pappa1993consistent} greater than
$0.9$.

\begin{table}[H]\centering
\begin{tabular}[c]{c| c| c| c| c| c}
\toprule
$ $ & $f_\text{SSI-cov} \text{[Hz]}$ & $\xi_\text{SSI-cov} [\%]$ & $f_\text{EFDD} \text{[Hz]}$ & $\xi_\text{EFDD} [\%]$ & $\text{MAC}_\text{SSI-ref,EFDD}$\\
\hline
\text{Mode 1}  & 2.68 & 3.47 & 2.69 & 2.97 & 0.99\\
\hline
\text{Mode 2}  & 3.37 & 3.90 & 3.35 & 4.11 & 0.99\\
\hline
\text{Mode 3}  & 6.21 & 1.44 & -- & -- & --\\
\hline
\text{Mode 4}  & 8.10 & 4.63 & 8.15 & 1.14 & 0.97\\
\hline
\text{Mode 5}  & 10.04 & 5.69 & 10.06 & -- & 0.97\\
\hline
\text{Mode 6}  & 11.95 & 1.15 & 12.24 & -- & 0.99\\
\bottomrule
\end{tabular}
\caption{Modal parameters of the tower, October 2017.}
\label{tab_freq_o}
\end{table}


The two first mode shapes are bending mode along the west-east direction
and north--south direction, respectively, while the third mode corresponds
to torsional movement of the tower and a deflection of the two lateral walls connected to
its south--west portion. The other experimental mode shapes
are more uncertain: the fourth one is likely a torsion mode shape mixed
with bending along north-east/south-west direction, and the
fifth and sixth are higher--order bending mode shapes.

\subsection{FE model updating}\label{subsec4-3}

\begin{figure}[H]\centering
\includegraphics[width=14cm,trim=25 25 0 25]{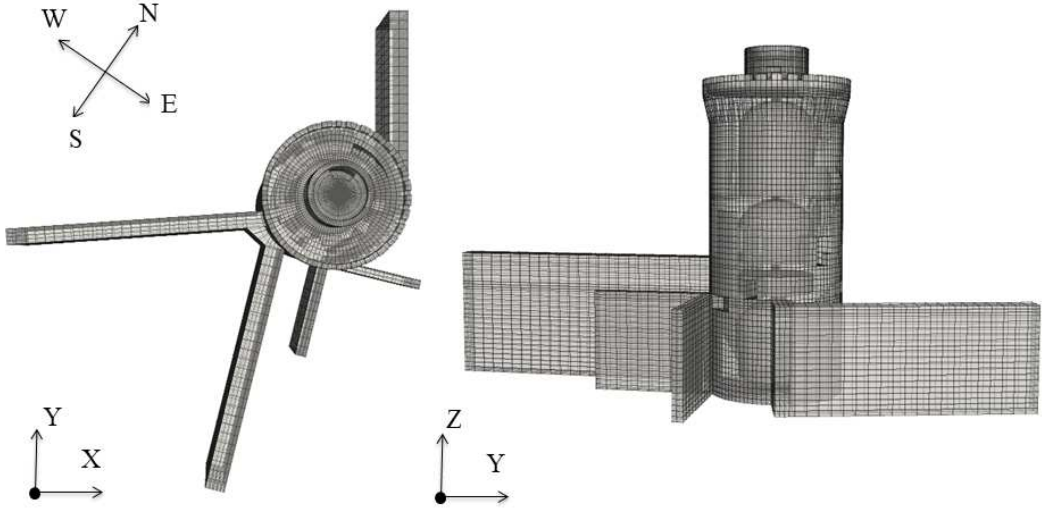}
\caption{FE model of the Matilde donjon.} \label{mastio_mesh}
\end{figure}

In this subsection, the procedure described in Section \ref{sec:numericalmethod} is
applied to the Matilde donjon. The FE mesh of the tower, shown in
Figure \ref{mastio_mesh}, consists of $52560$ isoparametric eight-node
brick elements and $64380$ nodes, for a total of 193140
degrees of freedom. The model, as shown in the Figure, includes a
portion of the surrounding walls.
The bases of the tower and lateral walls are fixed, and the ends of the walls are prevented
from moving along the X and Y directions.

\begin{figure}[H]\centering
\includegraphics[width=14cm,trim=25 25 0 25]{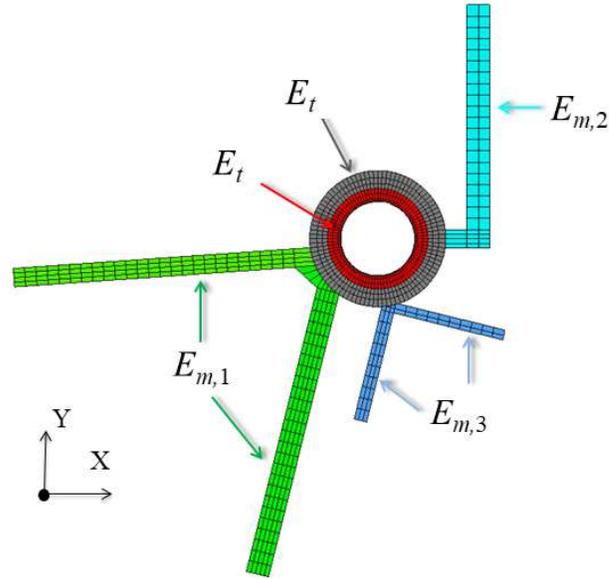}
\caption{Designated tower materials.} \label{mastio_mat}
\end{figure}

The numerical procedure has been used to estimate the values of the Young's modulus of the inner and outer layers ($E_{t,i}=E_{t,e}=E_{t}$) of the tower's walls, and Young's moduli ($E_{m,i}$) of the masonry constituting the Fortress' walls (Figure \ref{mastio_mat}), with $\mathbf{x}=[E_t, E_{m,1}, E_{m,2}, E_{m,3}]$. These parameters have been allowed to vary within the intervals \cite{circ2019}, \cite{CNRDT212}\\

\begin{equation}
\unit{1.00}{\giga\pascal}\leq\emph{$E_{t}$}\leq\unit{5.00}{\giga\pascal},
\end{equation}

\begin{equation}
\unit{1.00}{\giga\pascal}\leq\emph{$E_{m,1}$},\emph{$E_{m,2}$},\emph{$E_{m,3}$}\leq\unit{6.00}{\giga\pascal}.
\end{equation}

The Poisson's ratio of masonry is fixed at $0.2$, the mass density of the tower's walls is fixed at
$\unit{1800}{\kilogram\per\metre^3}$ and $\unit{2000}{\kilogram\per\metre^3}$
for the inner and outer layer, respectively, and the mass density of the side walls is taken to be
$\unit{2000}{\kilogram\per\metre^3}$. The experimental frequencies estimated by the SSI--cov method are
used in the optimization process, hence

\begin{eqnarray}\label{freqmastio}
\widehat{\mathbf{f}}=\unit{\text{[2.68, 3.37, 6.21, 8.10, 10.04, 11.95]}}{\hertz}.
\end{eqnarray}
The  optimal parameters are reported in table \ref{mastiopar}: the values of $\zeta$ and $\eta$  guarantee the reliability of $E_t$ and $E_{m,1}$, while the constituent materials the remaining walls are marked by uncertainty. The values obtained can be considered acceptable as the greatest uncertainty affects a part
of the structure, the right sidewall, whose geometric characteristics (thickness, height, composition),
connection degree with the tower and dynamic properties are unknown. Anyway, the optimal parameter values obtained can describe
the global dynamic behaviour of the tower.
 The total computation time for the model updating procedure
was 8468.9 s, and the number of evaluations $131$.


\begin{table}[H]\centering
\begin{tabular}[c]{c| c| c| c| c| c|}
\toprule
 & $x_j$ & $\zeta_j$ & $\eta_j$ & $ \zeta_j^{-1} $ & $ \eta_j^{-1} $\\
\hline
$E_{t} [\text{GPa}]$  & 2.152 & 1.627 & 1.557 & 0.615 & 0.642\\
\hline
$E_{m,1} [\text{GPa}]$  & 5.808 & 9.577$\cdot10^{-1}$ & 9.017$\cdot10^{-1}$ & 1.044 & 1.109\\
\hline
$E_{m,2} [\text{GPa}]$  & 5.532 & 6.409$\cdot10^{-2}$ & 1.139$\cdot10^{-2}$ & 15.603 & 71.942\\
\hline
$E_{m,3} [\text{GPa}]$  & 2.095 & 6.845$\cdot10^{-2}$ & 4.445$\cdot10^{-2}$ & 14.609 & 22.471\\
\bottomrule
\end{tabular}
\caption{Optimal parameter values calculated by NOSA--ITACA.} \label{mastiopar}
\end{table}

Table \ref{tab_freq_matilde} summarizes the numerical frequencies of the tower corresponding to the optimal parameters and their relative errors
$|\Delta_f|$ with respect to the experimental counterparts; $|\Delta_f|$
varies between $2$ and $3\%$, except for the third and sixth frequencies.


\begin{table}[H]
  \begin{center}
    \begin{tabular}{*{4}{c}}
    \toprule
     & $\widehat f_i \text{ [Hz]}$ & $f_i \text{ [Hz]}$ & {$|\Delta_f|  [\%]$} \\
    \hline
    mode 1 & $ 2.68 $ & $ 2.76 $ & $ 2.99 $ \\
    mode 2 & $ 3.37 $ & $ 3.33 $ & $ 1.19 $ \\
    mode 3 & $ 6.21 $ & $ 6.51 $ & $ 4.83 $ \\
    mode 4 & $ 8.10 $ & $ 7.90 $ & $ 2.47 $ \\
    mode 5 & $ 10.04 $ & $ 9.81 $ & $ 2.29 $ \\
    mode 6 & $ 11.95 $ & $ 11.10$ & $ 7.11 $ \\
    \bottomrule
    \end{tabular}
  \end{center}
\caption{Experimental frequencies
$\widehat {\mathbf f}$ and numerical frequencies $\mathbf f$ calculated for the optimal values of the parameters recovered by NOSA--ITACA.}\label{tab_freq_matilde}
\end{table}

\begin{figure}[H]\centering
\includegraphics[width=14cm,trim=25 25 0 25]{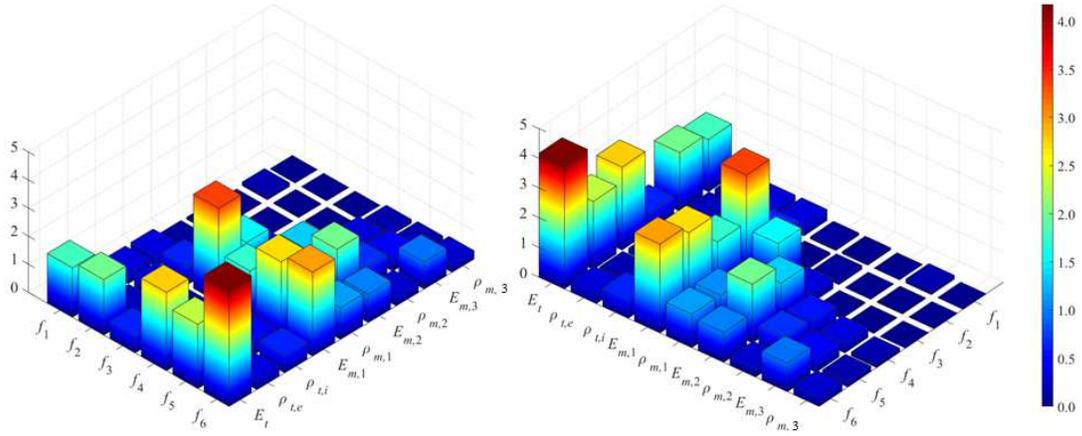}
\caption{EET sensitivity indices for the first sixth frequencies and nine parameters.} \label{EET_mastio}
\end{figure}

As for the simulated example, a GSA has been performed to validate the results of the sensitivity
analysis achieved by NOSA--ITACA. The EET method is used to evaluate the sensitivity
indices assuming a uniform probability distribution function, for the nine input factors (Young's modulus and mass density of each material),
and the Latin Hypercube as sampling
strategy; $500$ FE modal analyses were carried out.
Figure \ref{EET_mastio} shows that the elastic moduli of the tower and wall $1$
strongly influence the frequency variation as compared to the others. In particular, the tower's Young's modulus
impacts all frequencies except for the third, which is instead heavily affected by elastic modulus $E_{m,1}$,
as confirmed by the experimental mode shape which exhibits a large displacement component corresponding to an
out-of-plane deflection of the wall. The GSA analysis confirms the reliability of the NOSA--ITACA results.

\section{Conclusions}
\label{sec:conclusions}
The present paper proposes an improved numerical method to solve the constrained minimum problem encountered in FE model updating and calculate a global minimum point of the objective function in the feasible set. The global optimization method, consisting of a recursive procedure based on construction of local parametric reduced-order models embedded in a trust-region scheme, is integrated into the FE code NOSA-ITACA, a software developed in house by the authors. Along with the global optimization method, some issues related to the reliability of the recovered solution are presented and discussed.
In particular, once the optimal parameter vector has been calculated, two quantities involving the Jacobian of the numerical frequencies provide a measure of how trustworthy the single parameter is. The numerical method has been tested on two simulated examples, a masonry tower and a domed temple, in order to highlight the capabilities and features of the proposed global optimization algorithm. The results of the test cases, validated via a generic genetic algorithm and a global sensitivity analysis, prove the method's efficiency and robustness.
The objective function may have multiple local minimum points, and the first example highlights that the proposed procedure, unlike a genetic algorithm,  can provide a set of local minimum
points, including the global one. The second example shows some features of the code, which can help users to choose the most suitable optimal parameters characterized by higher reliability. Comparison of the computation time and number of objective function evaluations highlights that the NOSA-ITACA code performs better than the genetic algorithm. Regarding how the parameter variations can influence the frequencies of the FE model, the numerical method seems to provide the same information given by a global sensitivity analysis.
Finally, the paper has addressed a real case study the Matilde donjon in Livorno. The experimental dynamic properties of the historic tower monitored under operational conditions were used in the model updating procedure to estimate the mechanical properties of its constituent materials. The optimal parameter values obtained can describe
the global dynamic behaviour of the tower with a maximum error of $5\%$ on all the frequencies, except for the sixth. 

\section*{CRediT authorship contribution statement}
All authors listed have made a substantial, direct and intellectual
contribution to the work, and approved it for publication.

\section*{Acknowledgements}
The authors wish to thank Dr. Riccardo Mario Azzara, INGV Arezzo, for having made available the seismic instrumentation used in the experimental tests performed on the
Matilde donjon.

\bibliographystyle{unsrt}
\bibliography{paper_GFEMU}

\end{document}